# NETWORK EVOLUTION AND QOS PROVISIONING FOR INTEGRATED FEMTOCELL/MACROCELL NETWORKS

Mostafa Zaman Chowdhury[1], Yeong Min Jang[2], and Zygmunt J. Haas[3]

[1,2]Department of Electronics Engineering, *Wireless Networks and Communications Lab*
Kookmin University, Seoul, Korea
[1]mzceee@yahoo.com, [2]yjang@kookmin.ac.kr

[3] Department of Electrical and Computer Engineering, *Wireless Networks Lab*
Cornell University, Ithaca, NY, USA; haas@ece.cornell.edu

## ABSTRACT

*Integrated femtocell/macrocell networks, comprising a conventional cellular network overlaid with femtocells, offer an economically appealing way to improve coverage, quality of service, and access network capacity. The key element to successful femtocells/macrocell integration lies in its self-organizing capability. Provisioning of quality of service is the main technical challenge of the femtocell/macrocell integrated networks, while the main administrative challenge is the choice of the proper evolutionary path from the existing macrocellular networks to the integrated network. In this article, we introduce three integrated network architectures which, while increasing the access capacity, they also reduce the deployment and operational costs. Then, we discuss a number of technical issues, which are key to making such integration a reality, and we offer possible approaches to their solution. These issues include efficient frequency and interference management, quality of service provisioning of the xDSL-based backhaul networks, and intelligent handover control.*

## 1. INTRODUCTION

*Femto-access points (FAPs)* are low-power, small-size home-placed Base Stations (also known as *Home NodeB* or *Home eNodeB*) that create islands of increased capacity in addition to the capacity provided by the cellular system. These areas of increased capacity are referred to as *femtocells.* Femtocells operate in the spectrum licensed for cellular service providers. The key feature of the femtocell technology is that users require no new equipment (UE). However, to allow operation in the licensed cellular spectrum, coordination between the femtocells and the macrocell infrastructure is required. Such coordination is realized by having the femtocells connected to the local mobile operator's network using one or more of the following backhaul network technologies: residential xDSL, cable TV (CATV), Metro Ethernet, or WiMAX.

It is envisioned that a FAP would be typically designed to support simultaneous cellular access of two to six mobile users in residential or small indoor environments. Predictions show that in the near future about 60% of voice traffic and about 90% of data traffic will originate from indoor environments, such as a home, an office, an airport, and a school [1]. Therefore, there is the need for improved indoor coverage with larger data rates. However, due to the limited cellular capacity, it might be difficult and too expensive to accommodate this increased traffic demand using the current macrocellular coverage. Femtocells are a new candidate technology that is capable of providing expanded coverage with increased data rates.

Among the benefits of femtocell are low-cost deployment, reduced transmission power,

backward compatibility with the macrocellular technology, portability of devices, and scalable deployment [2]-[5]. Due to an interest from operators (such as the NGMN (Next Generation Mobile Network) Alliance) and standardization bodies (such as 3GPP, Femto Forum, Broadband Forum, 3GPP2, IEEE 802.16m, WiMAX Forum, GSMA, ITU-T, and ITU-R WP5D), integrated femtocell/macrocell is expected to be a major part of the IMT-Advanced network architecture [6]. From the wireless operator point of view, the most important advantage of the integrated femtocell/macrocell architecture is the ability to offload a large amount of traffic from the macrocell network to the femtocell network. This will not only reduce the investment capital, the maintenance expenses, and the operational costs, but will also improve the reliability of the cellular networks.

Even though there are several technical approaches to improve the indoor coverage, femtocell appears to be the most attractive alternative. Compared with the *Fixed Mobile Convergence (FMC)* framework, femtocells allow servicing large numbers of indoor users. In contrast with the 3G/Wi-Fi UMA(*Unlicensed Mobile Access*) technology, femtocell do not require dual-mode handset. Furthermore, another drawback of Wi-Fi is its use of the increasingly crowded unlicensed ISM band that causes significant interference. Finally, repeaters (or signal booster) [2] improve the wireless access coverage, but not the wireless capacity. Repeaters need new backhaul connections and only solve the poor coverage problem in remote areas, where fixed broadband penetration is low. The ABI Research [7] expects cellular-based femtocells to outpace UMA and SIP-based Wi-Fi solutions by 2013, grabbing 62 % of the market.

Fig. 1 shows an example of macrocellular network integration with femtocells [8]. Macrocells are operated by a mobile wireless operator, while femtocells are privately owned and connected to a broadband service provider, such as an Internet Service Provider (ISP). Thousands of femtocells may co-exist in a coverage area of a macrocell-based cellular network.

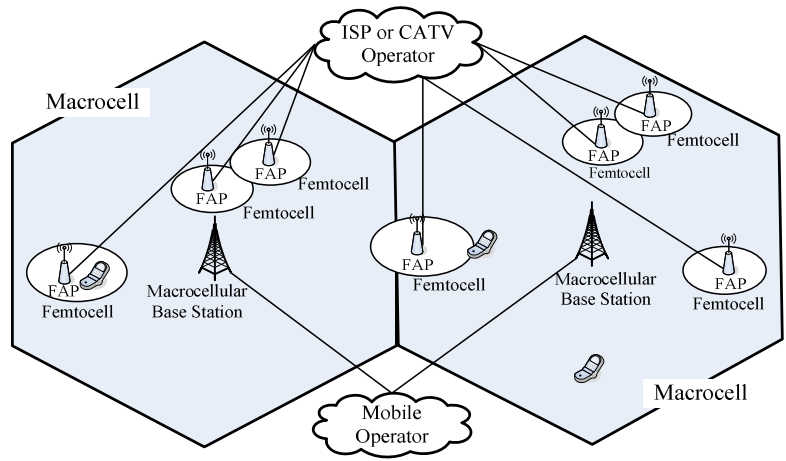

Figure 1: Integration of a macrocellular network with femtocells

The advantages of the femtocellular technology can be examined from multiple perspectives, such as the users, the manufacturers, the application developers, the network operators, and the service and content providers. For example, from the user's perspective, customers typically expect high data-rate wireless indoor access at low cost and with good quality of service (QoS). Of course, the key advantage of the femtocells for users is that there is no need for an expensive dual-mode handset; rather, the same single-mode handset is used to access the FAPs and the macrocellular network.

From the perspective of a network operator, the following are the key advantages of the integrated femtocell/macrocell network:

- **Improved coverage:** Providing extensive in-building coverage has long been a challenge for mobile operators. This is even a more difficult problem for communication at higher frequencies, where radio propagation loss is larger. Femtocells can provide better signal reception within the indoor environment, as FAPs are also located inside the building. Thus, using of the basic concept of spectrum re-use, femtocells can improve the network coverage and increase the network capacity. Especially, femtocells extend the service coverage into remote or indoor areas, where access to a macrocellular network is unavailable or is limited. Of course, improved coverage and access capacity enhances customer satisfaction, allowing the network operator to retain and expand its customer pool.

- **Reduced infrastructure and capital costs:** Femtocells use the existing home broadband connectivity for backhauling the femtocells' traffic. Thus, by steering users' traffic into their own FAPs and away from the macrocells, femtocells reduce the expensive backhaul costs of macrocellular networks.

- **Power saving:** Since femtocells target indoor coverage, the transmission power is significantly smaller, as compared to the transmission power of the macrocellular network that is required to penetrate into buildings. Smaller transmission power results in decreased battery drainage of the mobile devices, prolonging the devices' lifetime. Furthermore, decrease in the transmission power reduces inter-cell interference, increasing the signal-to-interference-ratio. This, in turn, improves the reception, increasing capacity and coverage.

- **Provisioning of QoS:** The radio path loss close to the fringe of a macrocell can be quite severe. Femtocells can in particular improve the QoS for users with poor macrocell reception.

Fig. 2 shows three possible types of femtocell network configurations based on the availability of a broadband connection (e.g., ISP) and on the coverage of the macrocellular network.

**Type A - a single stand-alone femtocell:** This could be the case of a remote area with no macrocellular coverage or a poor coverage area (such as an indoor or macrocell edge), and when no other neighboring femtocells are available. This type of a configuration extends the service coverage into remote areas.

**Type B - a network of stand-alone femtocells:** In this scenario, multiple FAPs are situated within an area in such a way that a radio signal from one FAP overlaps with other FAPs' signals. There is no macrocellular coverage or the coverage is poor. Femtocell-to-femtocell handovers are present and need to be handled by the femtocellular network. As the Type A configuration, the Type B configuration is also able to extend the service coverage into remote areas.

**Type C - a femtocell network integrated with a macrocellular infrastructure:** This scenario can be viewed as a two-tier hierarchical network, where the macrocells create the upper tier and the femtocells the lower tier. Handover between macrocells and femtocells, as well as handover between femtocells, are common occurrence in this scenario. This configuration improves the indoor service quality and reduces the traffic load of the macrocells by diverting traffic to femtocells.

There are several technical, business, and regulatory issues of the femtocellular technology that remain to be addressed [4], [5], [9]-[11]. As the femtocellular network co-exists with the macrocellular infrastructure, the technical issues relate to radio resource management, end-to-end QoS support for network architecture, and mobility management. In particular, the main technical issues that affect the performance of the integrated network and influence the QoS level of the femtocell users are: frequency and interference management, service provisioning in capacity limited xDSL backhaul, and handover between a macrocell and femtocells.

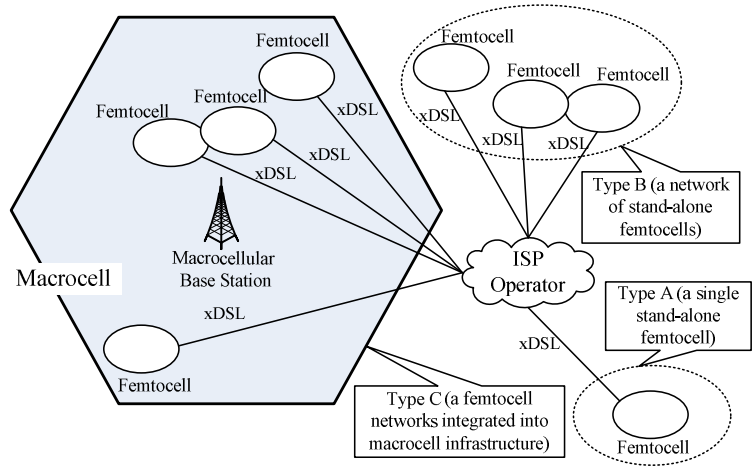

Figure 2: The three types of the femtocellular network configurations

The goal of this paper is to propose an evolutionary path from the existing macrocellular networks into the integrated femtocell/macrocell network architecture and to suggest solutions to the QoS provisioning of the integrated network.

## 2. THE EVOLUTION OF NETWORK ARCHITECTURE TO INTEGRATED FEMTOCELL/MACROCELL NETWORKS

In the proposed three types of the femtocell deployment in Fig. 2, the number of FAPs, their locations, and identification parameters change dynamically. Integration with the core network (CN) should be possible using existing standardized interfaces. Some issues that should be considered for the network integration are low capital expenditure (CAPEX) and low operational expenditure (OPEX), coexistence with other wireless networks, higher spectral efficiency, improved cell-edge performance, scalability of the provisioning and the management processes, self-organizing network (SON) architecture, and QoS support. Fig. 3 depicts the basic device-to-CN connectivity for the femtocell network deployment [3], [8], [12], [13]. The femtocell access networks are connected to CN through backhaul networks. A femtocell management system (FMS) is used to control and to manage the FAPs within a regional area. The FMS functionalities include configuration, fault detection, fault management, monitoring, and software upgrades.

There are several deployment options for the femtocell/macrocell network integration. For a particular situation, the choice of architecture depends on numerous factors, such as the size of the network, the existing network features, the future rollout and convergence plans of the network operator, capacity planning, the ability to co-exist and interact with the existing network, and the predicted network evolution. We propose three methods for integration of the femtocells with the macrocellular infrastructure:

- **Small-scale** deployment using the existing macrocell *Radio Network Controller* (RNC) interface
- **Medium- and large-scale** deployment using a concentrator and *IP Multimedia Subsystem (IMS)*
- **Large-scale and highly-dense** integration using *SON* and cognitive radio *(CR)* along with concentrator and *IMS*

The "small-scale," "medium- and large-scale," and "large-scale and highly-dense" femtocell deployment network architectures differ in terms of network entities, connecting procedures, and management systems. The size of the scale refers to the number and density of FAPs that are to be installed. For small-scale femtocell deployment, to reduce the implementation costs, the existing macrocellular RNC interface is used for FAPs connection to the macrocellular infrastructure. For medium- and large-scale femtocell deployment, the existing network infrastructure is insufficient to support the number of FAPs, and introduction of new communication elements, or modification of the existing infrastructure is necessary [4]. One of such new elements is the *Femto-GateWay* (*FGW*), which is able to control many FAPs. IMS provides an efficient way to control a large number of signaling messages using the *Session Initiated Protocol* (*SIP*). As the number of FAPs within an area increases, the interference from neighbor FAPs increases as well, as does the rate of handovers. This is where SON [14] architectures can be useful in reducing the interference by the auto-configuration of frequency allocations and by self-adapting the transmission power of the neighbor FAPs. A SON architecture is also able to improve handover performance.

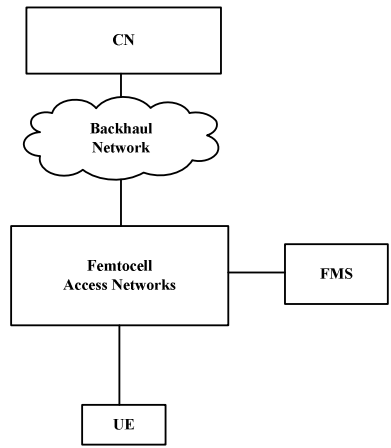

Figure 3: The basic femtocellular network architecture

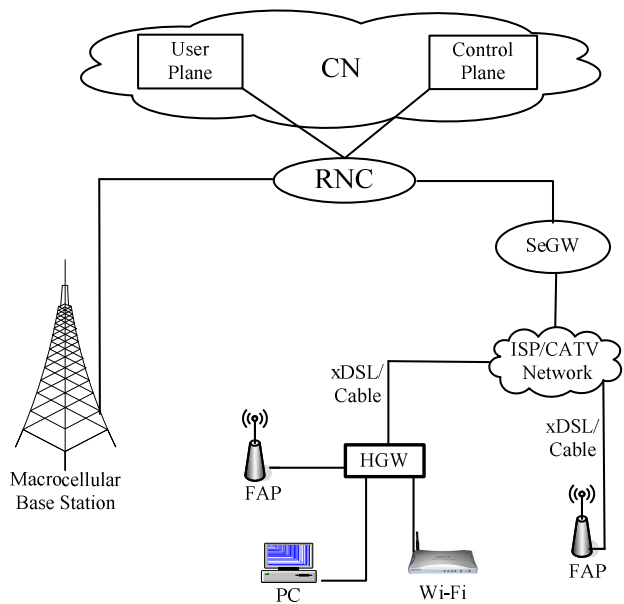

Figure 4: The proposed small-scale deployment of the integrated femtocell/macrocell network architecture using the existing macrocellular RNC interface

## 2.1. The Small-Scale Deployment using the Existing Macrocellular RNC Interface

Fig. 4 shows the femtocell/macrocell integrated network architecture for small-scale deployment. Such architecture could be useful in a rural or a remote area, where only a few femtocells are required.

This architecture is similar in its control part to the existing UMTS-based 3G network; each FAP is corresponding to a macrocellular Base Station and is connected to the RNC. The *Security GateWay* (*SeGW*) implements a secure communication tunnel between FAP and RNC, providing mutual authentication, encryption, and data integrity functions for signalling, voice, and data traffic [8]. Security of communications between the FAP and the *SeGW* can also be handled by the *IPsec* protocol. The RNC is responsible for controlling and managing the macrocellular Base Stations, together with the newly added FAPs. The pros of this architecture are that it is simple and cost-effective for a small number of FAPs and that it uses the existing RNC interface. The disadvantages of the architecture are that it is not scalable to a larger number of FAPs and that the RNC capacity to support the macrocellular infrastructure is reduced.

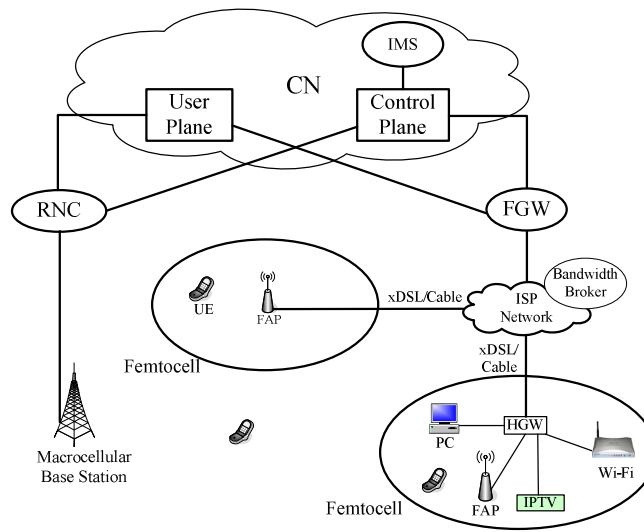

Figure 5: The proposed medium- and large-scale deployment of the integrated femtocell/macrocell network architecture using concentrator and IMS

## 2.2. The Medium- and Large-Scale Deployment using a Concentrator and IMS

HSDPA/HSUPA cellular networks utilize centralized RNC to control their associated Base Stations. A single RNC is in charge of radio resource management of approximately 100 macrocellular Base Stations. However, since it is envisioned that thousands of femtocells may exist within an area of a single macrocell, a single RNC is incapable of controlling such a large number of femtocells. Therefore, a different and more efficient femtocell control scheme is necessary. One such scheme in Fig. 5 implements a concentrator-based UE-to-CN connectivity for integrated femtocell/macrocell networks [4], [5], [8]. The FGW acts as a concentrator and several FAPs are connected to FGW through a broadband network, such as an ISP, for example. There is no direct interface between the RNC and FGW, so the FGW communicates with RNC through CN. The FGW manages thousands of femtocells and appears as a legacy RNC to the existing CN. Inter-operability between the mobile operator and the ISP network and among the mobile operators is required to connect the femtocell users to other users. Whenever a FAP is installed, a mobile operator stores its location information gathered from the macrocellular networks using femtocell searching (sniffing) for management purpose.

Other alternatives to find location information are using the customer contractual billing address information of the ISP operator and the GPS (Global Positioning System) module, which could be optionally implemented on a FAP, but is likely to suffer from poor satellite coverage.

The main advantages of IMS-supported integrated femtocell/macrocell networks are scalability and the possibility of rapid standardization ([9]). IMS also supports seamless mobility, which is required by cellular operators [15]. A policy and an IMS-based *Bandwidth Broker (BB)* [16] is required to control the end-to-end bandwidth allocation. The advantages of this architecture are that it supports a large number of FAPs, that it supports better QoS than the small-scale deployment architecture, and that the RNC capacity is not affected by FAPs' deployment. The disadvantage of this architecture is that it is not cost effective for a small number of FAPs.

## 2.3. The Large-Scale and Highly-Dense Deployment Architecture using SON and CR along with Concentrator and IMS

Traditionally, macrocellular networks require complex and expensive manual planning and configuration. Currently, 3GPP LTE-Advanced and IEEE 802.16m are standardizing the SON concept for IMT-Advanced networks. The main functionalities of SON for integrated femtocell/macrocell networks are self-configuration, self-optimization, and self-healing [14]. The self-configuration function includes intelligent frequency allocation among neighboring FAPs; self-optimization attribute includes optimization of transmission power among neighboring FAPs, maintenance of neighbor cell list, coverage control, and robust mobility management; and self-healing feature includes automatic detection and resolution of most failures. The sniffing function is required to integrate femtocell into a macrocellular network, so that a FAP can scan the air interface for available frequencies and other network resources. Self-organization of radio network access is regarded as a new approach that reduces OPEX/CAPEX. It enables cost-effective support of a range of high-quality mobile communication services and applications for affordable prices in a dense femtocell deployment. An advanced self-organizing mechanism enables deployment of dense femtocell clusters [17] and the integrated femtocell/macrocell networks should incorporate SON capabilities. This way, neighboring FAPs can communicate with each other to reconfigure resources, transmission powers, and frequency assignments. Network operators may need to deploy the hierarchical femtocell network architecture based on centralized or distributed SON. The centralized SON architecture may be necessary for cooperative hotspot coverage, but the distributed and flat SON architecture may be required for individual, ad-hoc, and random femtocell coverage. Therefore, further enhancements of SON will be an important element of a future femtocell deployment. The ultimate deployment architectures of the IMT-Advanced network will rely on concentrator-based IMS/SIP and SON capable all-IP network architecture. The future FAPs may also support personal ubiquitous home networking services to control home devices and machine-to-machine communications. Fig. 6 shows the basic features and framework of the proposed SON-capable integrated femtocell/macrocell network architecture of the large-scale and highly-dense femtocell network. Next, we briefly discuss three examples of operations of a femtocell network.

**Scenario A - frequency configuration and power optimization:** If large numbers of FAPs are deployed in an indoor building or femto-zone area, signals from different FAPs will interfere with each other. The FAPs need to coordinate with each other to configure frequency and optimize transmission power.

**Scenario B - interference mitigation and cell size adjustment:** If an UE connected to a FAP that is designated as a master FAP receives interfering signal from other FAPs, then the master FAP requests these FAPs to reconfigure their transmission power, so that interference at the UE is reduced. The neighboring femtocells re-adjust their cell size to reduce the interference.

**Scenario C - a seamless handover:** When an UE moves from a macrocell to a femtocell or from a femtocell to another femtocell and detects more than one FAP, the FAPs and macrocellular Base Station (e.g., *eNodeB*) coordinate with each other to facilitate fast and seamless handover.

The advantages of this architecture are that it supports large number and high density of FAPs, that it supports better QoS than the other two architectures, that it can reduce interference effects, and that it improves spectral efficiency. The disadvantage of this architecture is that it is not economically suitable for a deployment of a small number of FAPs.

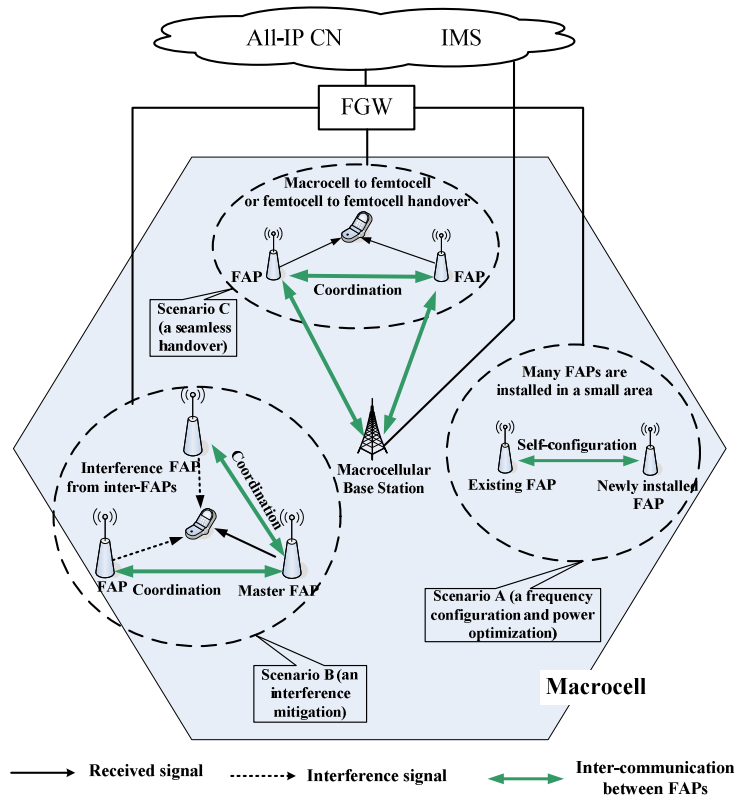

Figure 6: The proposed SON features for the large-scale and highly-dense deployment of an integrated femtocell/macrocell network architecture

Future all-IP CN will provide a common IP-based network platform for heterogeneous access networks, advanced mobility management, enhanced session management, and network extension/composition [18]. These features will enhance the performance of the integrated femtocell/macrocell networks.

## 3. THE TECHNICAL CHALLENGES OF QOS PROVISIONING IN FEMTOCELL/MACROCELL INTEGRATED NETWORKS

The provision of QoS in femtocellular networks is more difficult than for the existing macrocellular networks due to the large number of neighboring FAPs and the possible interference conditions among the femtocells and between macrocells and femtocells [19]. There are a number of technical challenges to support satisfactory QoS to the users in a femtocellular network, solutions of which are being studied. The QoS of femtocellular networks is influenced by procedures such as resource allocation, network architecture, frequency and interference management, power control, handover control, security assurance, and QoS provisioning in backhaul networks. From among the many QoS issues, we briefly discuss only three: frequency and interference management, QoS provision in xDSL-based backhaul, and handover control.

### 3.1. Frequency and Interference Management

To support QoS and increased capacity in integrated macrocell/femtocell networks, dynamic frequency and interference management, similar to those in macrocellular networks, is necessary. At the femtocell edges, users experience significantly more interference than users located closer to the FAP [9]. Intelligent and automated radio-frequency planning is needed to minimize interference for random and unknown installation of femtocells. In 3GPP [3], a number of different deployment configurations have been considered for FAP. Dedicating a channel is possible for femtocell users operating in an unused frequency band of the macrocells. Since in this approach there is no interference between femtocell users and macrocell users, it is an optimal solution that avoids the interference problem altogether. But this solution also reduces the spectral efficiency, so it is quite expensive and, thus, undesirable. Hence, sharing frequency bands between femtocells and macrocells seems to be a more appropriate, but also technically more challenging, solution.

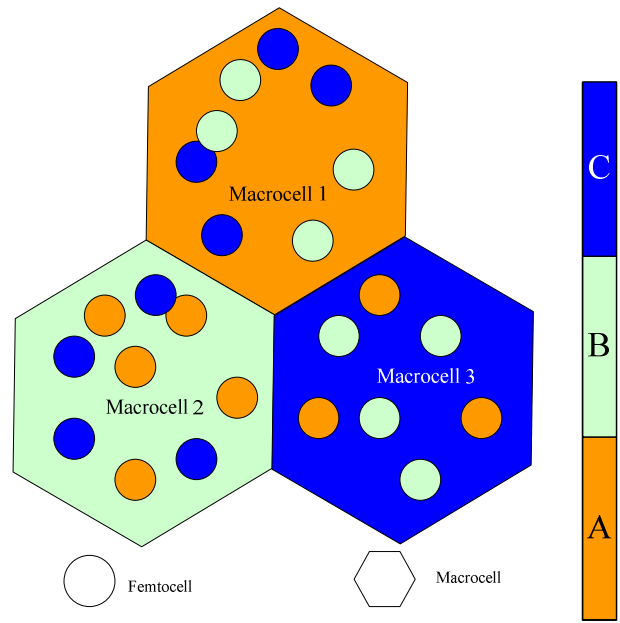

Figure 7: A possible frequency allocation among femtocells and macrocell to reduce interference

In the models of Figs. 4 and 5, for easy deployment and management of FAPs, the conventional re-use scheme could be employed, where the frequency re-use factor is fixed throughout the macrocellular network and the femtocellular networks. However, frequency re-use should be carefully managed, as improper frequency allocation among femtocells could cause significant co-channel interference. For the model with a small number of FAPs (Fig. 4), the solution of sharing a frequency band by femtocells with the macrocellular network seems to be appropriate because the number of FAPs is too small to create significant macrocell-to-femtocell or inter-femtocell interference. However, for the model with a large number of FAPs (Fig. 5), sharing a frequency band may not be a suitable solution, because of the large femtocell-to-macrocell interference. In this latter case, the traditional re-use scheme could be implemented. Fig. 7 shows an example of such a frequency allocation scheme with re-use factor of 3, where the frequency band is divided into three sub-bands (*A, B,* and *C*), and each sub-band allocated to one macrocell of three-macrocell cluster. The femtocells within a macrocell use one of the other two sub-bands. For instance, assume that the macrocell 1 uses sub-band *A*, then the FAPs within this macrocell should use sub-bands *B* or *C* to avoid femtocell-to-macrocell interference. Approximately half of the FAPs within a macrocell should use each of the other sub-bands, as to minimize the inter-femtocell interference. Moreover, interference mitigation techniques, such as power control or directional antennas, implemented through a SON-based architecture could be used to further reduce the inter-femtocell interference.

Fig. 8 shows the outage probability comparison when 1000 random femtocells are placed within a macrocell. Other assumptions used in this outage probability analysis are the same as in [11]. The proposed frequency allocation scheme is able to reduce the outage probability, as compared with the scheme which uses the same frequency band allocation (i.e., the same frequency band is allocated for all the femtocells and the overlaid macrocell) and with the scheme with dedicated frequency allocation (i.e., one frequency band is shared by all the femtocells and a different frequency band is allocated to the overlaid macrocell). The same frequency band allocation scheme creates large femtocell-to-femtocell interference as well as large femtocell-to-macrocell interference. On the other hand, the dedicated frequency band scheme creates large inter-femtocell interference, effectively decreasing the frequency re-uses factor.

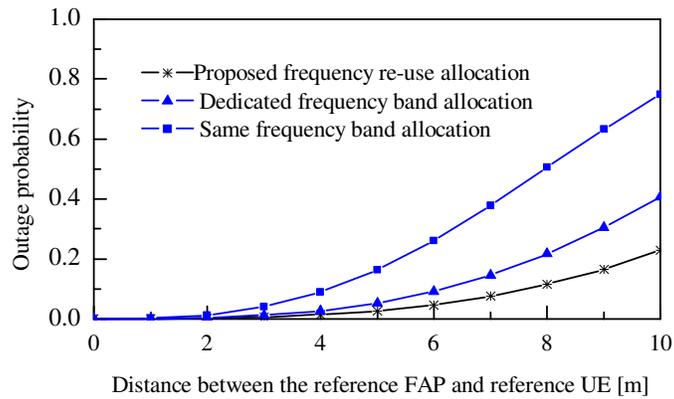

Figure 8: A comparison of outage probabilities

As the density of FAPs increases, the chance of assigning the same frequency band to two neighboring FAPs femtocells increases, and the inter-FAPs' interference becomes more significant [20]. In the model in Fig. 6, LTE-Advanced and IMT-Advanced systems will support advanced interference management and mitigation. Also, schemes such as enhanced dynamic frequency re-use schemes for reduction of femtocell-to-macrocell interference and inter-femtocell interference in a dense FAPs area could be applied. To reduce the interference and for effective use of the spectrum, fractional frequency re-use (FFR) [11], [21], [22] is needed in the large- and densely-deployed femtocell scenario. FFR will be possible in the SON-based femtocell architecture, where it will minimize the interference by intelligent frequency allocation and power optimization. When a new FAP is installed, the frequency allocation among neighboring FAPs will auto-reconfigure, so that the frequencies used by near FAPs are different, thus reducing the inter-femtocell interference. The femtocell network will be characterized by highly efficient spectral utilization through adopting the autonomic and coordinated radio resource management technologies, such as cognitive radio, adaptive array antenna, cooperative SON technology, inter-frequency scheduling, synchronization control, power control, multiple diversity techniques, interference-based load balancing, and cooperative MAC and routing schemes.

### 3.2. QoS Provisioning in xDSL-based Backhaul

In a typical home environment, multiple pieces of equipment share the same xDSL connection by using a *Home Gateway (HGW),* a home networking device used to connect several pieces of equipment to the Internet. When femtocells are deployed using non-dedicated fixed broadband technology such as xDSL, preservation and maintenance of QoS require special attention [23]. Good voice quality requires low latency and small packet dropping probability, but none of these are guaranteed by the current ISP networks. Quality of the backhaul connection typically depends on the xDSL capacity, overall backhaul network load, congestion control, and bandwidth management. The DSL network does not control QoS on a call-by-call basis. The *HGW* or the routing gateway should classify data traffic by DiffServ or as best-effort traffic and should discriminate traffic by its type when the traffic leaves for the xDSL network. The class-based discriminated-service control rule should be set up in the *HGW* at configuration time [24].

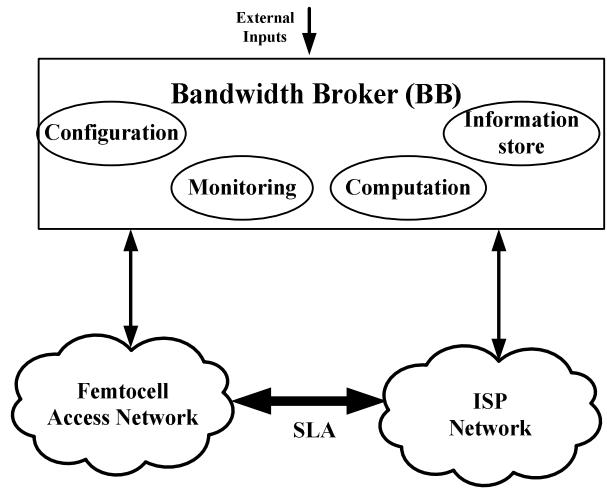

Figure 9: The proposed SLA framework to ensure QoS for the femtocell users

If no priority scheme is implemented, the small-scale deployment network architecture shown in Fig. 4 cannot guarantee the QoS for real-time traffic of femtocell users. Furthermore, for the model shown in Fig. 4, there is no *Service Level Agreement (SLA)* [10] or tight inter-operability between operators. The femtocell industry has an agreement with the Broadband Forum to use the TR-069 specifications [25]. This specification covers auto-configuration and dynamic service activation, remote management, firmware management, and performance monitoring. For successful femtocell deployment, ISP networks should have the capability of prioritizing suitably marked traffic. The QoS problem is reduced in the medium- and large-scale deployment network architecture, shown in Fig. 5 by introducing *SLA,* and *Bandwidth Broker (BB).* To meet the demands of today for applications such as IPTV and VoIP, some operators are considering the TR-098 specification [26] for remote QoS and service differentiation management for real-time applications. When *HGW* is used, the *Weighted Fair Queuing (WFQ),* a flow-based queuing algorithm, can also prioritize low volume femtocell voice traffic. An *SLA* between ISP and the mobile operator would benefit both operators and the QoS level of femtocell users would be ensured [10]. The *BB* concept may be used between the mobile operator and the ISP operator. A dynamic bandwidth reservation scheme [27], soft-QoS scheme [28], or other bandwidth management scheme could also be used to provide sufficient bandwidth to the femtocell users. The SON-based architecture, shown in Fig. 6, provides the advanced features for direct coordination and cooperative load balancing among neighboring FAPs. If xDSL capacity is limited, cooperative communication among neighboring FAPs could be used to divert traffic to other FAPs. Thus, the SON-based integrated network architecture would allow QoS guarantees. In Fig. 9, a framework to provide improved QoS for the femtocell users is shown.

In Fig. 9, the SLA between the ISP network and the femtocell access network ensures sufficient bandwidth for the femtocell users. The BB in this framework has four main functionalities: configuration, monitoring, computation, and database information storage. The BB is the central logical entity that is responsible for the execution of the SLA negotiation and resource allocation. The BB monitors the services and their bandwidth allocation by examining parameters such as the requested bandwidth and the type of services of the client networks (the ISP network and the femtocell network). The ISP installs a monitoring scheme to measure the bandwidth allocation among different access networks. The Database stores the information of the client networks, the SLA policy, and the connection history. The BB implements the policy of bandwidth allocation according to the monitoring information and based on specific set of rules. The amount of reserved bandwidth for femtocell users is calculated according to the configuration, the monitoring information, and the database information. The allocation of bandwidth could be controlled by a rate-control scheme that allows bandwidth reservation based on the application type.

### 3.3. Handover Control

The ability to seamlessly handover between the femtocellular and the macrocellular networks is a key element of the integrated architecture. Make-before-break handover may be necessary for the femtocell environments, because handovers will occur frequently due to the small size of the femtocells. As the coverage areas of femtocell may not be contiguous, there are four possible scenarios for handovers in the femtocell/macrocell integrated networks: macrocell-to-macrocell, macrocell-to-femtocell, femtocell-to-macrocell, and femtocell-to-femtocell. Delay, jitter, and packet loss are the main QoS parameters to consider for evaluation of a handover procedure. For the network architecture in Fig. 4, the handover effect is small due to the small number of neighboring FAPs. Thus, the existing macrocellular handover call flow for macrocell-femtocell handovers could be used. For the HSDPA/HSUPA network architecture in Fig. 5, there may be a need for a new handover call flow for macrocell-femtocell handover and for inter-femtocell handover, as well as for a new concentrator-based architecture and for IMS signaling. For the architecture in Fig. 6, the handover call flow procedure may be more involved and may include interference mitigation and the need to obtain the list of neighboring femtocell.

We are going to propose here a basic architecture of the macrocell-to-femtocell handover procedures for the model in Fig. 5. The handover from a macrocell to a femtocell is a challenging problem due to the fact that every macrocell may have thousands of neighboring femtocells. Identifying the neighboring FAPs and determining the optimal FAP as the handover target may be a complex decision-making process. Possible introduction of additional interference should also be considered as part of a handover decision. Fig. 10 shows the basic procedures of macrocell-to-femtocell handover for the concentrator-based integrated femtocell/macrocell network architecture. Although this scheme supports seamless handover, the UEs power consumption increases due to the need to scan many FAPs and, thus, the MAC overhead becomes significant. Furthermore, the increased size of the message carrying the neighbor FAP list results in increased overhead of the broadcasting operation. Thus, minimization of the neighbor FAP list is highly desirable.

The handover scheme shown in Fig. 10 provides an efficient mechanism to reduce the number of scanning. Due to fewer scans, fast handover is possible. Furthermore, fewer scans lead to a reduction in power consumption. Whenever the UE receives signals from multiple FAPs,[1] it relays the IDs of the detected FAPs to the macrocellular Base Station (steps 1, 2). As the Registration data base (DB) stores FAPs location, neighbor FAP list, and authorized FAP list, it provides the authorized neighbor FAP list to the UE (steps 3, 4). After receiving the authorized FAP list, the UE scans only the authorized FAPs and selects the best FAP for handover based on the measured value of the signal-to-interference ratio (e.g., the $E_b/I_0$ parameter) and on the direction and the speed of the UE (step 5). The UE sends a handover request message to the macrocellular Base Station, including the target FAP's ID, current $E_b/I_0$ value, scrambling code used for the target FAP, up/down link frequency of the target FAP, location code, routing code, and service area code (step 6(a)). This request is passed on to the target FAP through RNC, CN, and FGW (steps 6(b) to 6(e)). The call admission control (CAC) and radio resource control (RRC) at the FAP take into consideration the interference level and the availability of resources (xDSL bandwidth and radio interfaces) (step 7). The UE receives a response to its handover request, and it prepares for the physical channel reconfiguration (step 8). A radio link is then established between the UE and the FAP (step 9). After establishing a new link between the UE and FAP, the macrocellular Base Station tears down the connection with the UE (step 10).

---

[1] In our example, we assume that the UE receives from the target FAP, the neighbor FAP #1, and the neighbor FAP #2.

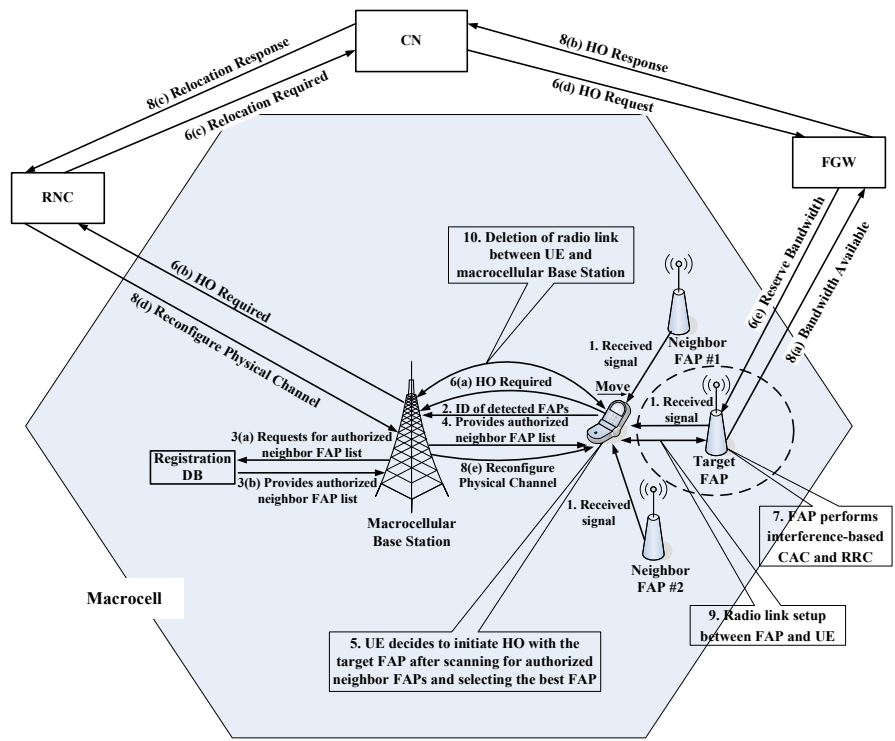

Figure 10: The basic call-flow procedures of the macrocell-to-femtocell handover for the concentrator-based femtocell network architecture

Frequent and unnecessary handovers that degrade the end-to-end QoS is a serious problem that needs to be considered [4]. The femtocell coverage area is very small, and there is a large chance that a fast moving UE will stay for a very short time within the coverage of a femtocell. Two special parameters, threshold time and threshold velocity, can be introduced in the CAC to minimize the unnecessary handovers.

The proposed handover scheme is also applicable to other advanced femtocell networks, such as the LTE-Advanced and the IMT-Advanced networks. The future SON-based architecture will provide an optimized neighbor FAP list, especially for dense femtocell deployment. The optimization of neighbor FAP list will provide a minimal set of FAPs necessary for handover.

The CR and SON-based architecture will be capable of detecting the accurate movement direction and speed of UEs required for the implementation of future handover schemes.

## 4. SUMMARY AND CONCLUDING NOTES

Integration of femtocellular networks with macrocellular networks, as well as with other non-wireless networks, is essential for the successful deployment of the femtocell technology. We proposed here three possible architectures for the evolution of the integrated network architecture, which are QoS-based and which support cost-effective integration. We discussed a number of the technical issues that need to be addressed to make such integration a reality and we offer possible approaches to solutions of these issues: efficient frequency and interference management, dynamic QoS provisioning of the xDSL-based backhaul, and intelligent handover control. The concentrator-based FAPs' deployment can support large number of FAPs, but cannot support QoS for dense FAPs' deployment. Frequency allocation, interference mitigation, and handover control for the dense deployment of FAPs are very complex procedures. Thus, an intelligent and automatic control system with a concentrator is needed to maximize the utilization of frequency bands, to minimize of femtocell-macrocell and inter-femtocell interference, and to ensure seamless and fast handover. The SON-based coordination and cooperative communications among FAPs and macrocellular Base Stations can improve the spectral utilization and QoS performances. Such communications and message sharing can reduce the number of scans, allowing fast handovers. It can also reduce the interference by optimizing transmission power and configuring frequency spectrum dynamically. An *SLA* between an ISP operator and a mobile operator can ensure bandwidth availability of limited backhaul networks for femtocell users. Consequently, in the near future, the integrated femtocell/macrocell networks should be equipped with self-optimizing, auto-configuring, interworking, and inter-operability capabilities to meet the requirement of IMT-Advanced. Future research should concentrate on economic feasibility of the femtocell/macrocell integrated network and on QoS-guaranteed mechanisms for the integrated network architecture, including movable and portable FAPs, to improve cell coverage, cell edge performance, UE's movement detection, FAP's location estimation, and support of fast handover in dense areas.

### ACKNOWLEDGEMENTS

This work was supported by the IT R&D program of MKE/KEIT [10035362, Development of Home Network Technology based on LED-ID].

### REFERENCES

[1] D. Lopez-Perez, G. de la Roche, A. Valcarce, A. Juttner, and J. Zhang, "Interference Avoidance and Dynamic Frequency Planning for WiMAX Femtocells Networks," *Proc. of IEEE Int. Conf. on Commun. Systems (ICS),* Nov. 2008.

[2] The Femto Forum, "Regulatory Aspects of Femtocells," May 2008.

[3] 3GPP TR R25.820, "3G Home NodeB Study Item," March 2008.

[4] M. Z. Chowdhury, W. Ryu, E. Rhee, and Y. M. Jang, "Handover between Macrocell and Femtocell for UMTS based Networks," Proc. of IEEE Int. Conf. on Adv. Commun. Technology (ICACT), Feb. 2009.

[5] H. Claussen, L. T. W. Ho, and L. G. Samuel, "An Overview of the Femtocell Concept," Bell Labs Technical Journal, 2008.

[6] L. Eastwood, S. Migaldi, Q. Xie, and V. Gupta, "Mobility Using IEEE 802.21 in a Heterogeneous IEEE 802.16/802.11-based, IMT-Advanced (4G) Network," IEEE Wireless Commun., April 2008.

[7] http://www.cellular-news.com/story/33163.php

[8] 3GPP TR R3.020, "Home (e)NodeB; Network Aspects," Sept. 2008.

[9] V. Chandrasekhar, J. G. Andrews, and A. Gatherer, "Femtocell Networks: A Survey," IEEE Commun. Magazine, Sept. 2008.

[10] M. Z. Chowdhury, S. W. Choi, Y. M. Jang, K. S. Park, and G. Yoo, "Dynamic SLA Negotiation using Bandwidth Broker for Femtocell Networks," IEEE Int. Conf. on Ubiquitous and Future Networks (ICUFN), June 2009.

[11] M. Z. Chowdhury, Y. M. Jang, and Z. J. Haas, "Interference Mitigation Using Dynamic Frequency Re-use for Dense Femtocell Network Architectures," IEEE Int. Conf. on Ubiquitous and Future Networks (ICUFN), June 2010.

[12] http://www.femtoforum.org

[13] http://www.airwalkcom.com

[14] 3GPP TS 32.500, "Telecommunication Management; Self-Organizing Networks (SON); Concepts and Requirements," Dec. 2009.

[15] P. Agrawal, J. H. Yeh, J. C. Chen, and T. Zhang, "IP Multimedia Subsystems in 3GPP and 3GPP2: Overview and Scalability Issues," IEEE Commun. Magazine, Jan. 2008.

[16] Z. Duan, Z. L. Zhang, Y. T. Hou, and L. Gao, "A Core Stateless Bandwidth Broker Architecture for Scalable Support of Guaranteed Services," IEEE Trans. on Parallel and Distributed Systems, Feb. 2004.

[17] H. Claussen, L. T. W. Ho, and L. G. Samuel, "Self-optimization of Coverage for Femtocell Deployment," WTS2008.

[18] 3GPP TR 22.978, "All-IP Network (AIPN) Feasibility Study," Dec. 2008.

[19] H. S. Jo, C. Mun, J. Moon, and J. G. Yook, ember, "Interference Mitigation Using Uplink Power Control for Two-Tier Femtocell Networks," IEEE Trans. on Wireless Commun., Oct. 2009.

[20] M. C. Necker, "Interference Coordination in Cellular OFDMA Network," IEEE Network, Nov./Dec. 2008.

[21] G. Boudreau, J. Panicker, N. Guo, R. Chang, N.g Wang, and S. Vrzic, "Interference Coordination and Cancellation for 4G Networks," IEEE Commun. Magazine, April 2009.

[22] S. H. Ali, and V. C. M. Leung, "Dynamic Frequency Allocation in Fractional Frequency Reused OFDMA Networks" IEEE Trans. on Wireless Commun, Aug. 2009.

[23] D. N. Knisely, T. Yoshizawa, and F. Favichia, "Standardization of Femtocells in 3GPP," IEEE Commun. Magazine, Sept. 2009.

[24] J. Song, M. Y. Chang, and S. S. Lee, and J. Joung, "Overview of ITU-T NGN QoS Control," IEEE Commun. Magazine, Sept. 2007.

[25] DSL Forum TR-069, "CPE WAN Management Protocol," May 2004.

[26] Broadband Forum TR-098, "Internet Gateway Device Data Model," 2007.

[27] X. Chen, B. Li, and Y. Fang, "A Dynamic Multiple-Threshold Bandwidth Reservation (DMTBR) Scheme for QoS Provisioning in Multimedia Wireless Networks," IEEE Trans. on Wireless Commun., March 2005.

[28] C. W. Leong and W. Zhuang, "Soft QoS in Call Admission Control for Wireless Personal Communications," Wireless Personal Commun., 2002.

**Mostafa Zaman Chowdhury** received his B.Sc. in Electrical and Electronic Engineering from Khulna University of Engineering and Technology (KUET), Bangladesh in 2002. Then, he joined as a faculty member the Electrical and Electronic Engineering department of KUET, Bangladesh in 2003. He completed his MS from Wireless Networks and Communications Lab. of Kookmin University, Korea in 2008. Currently, he is continuing his Ph.D. studies in Wireless Networks and Communications Lab. of Kookmin University, Korea. His current research interests focus on convergence networks, QoS provisioning, mobility management, and femtocell networks.

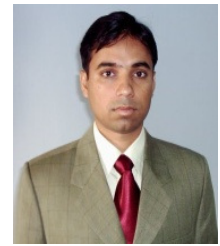

**Yeong Min Jang** received the B.E. and M.E. degree in Electronics Engineering from Kyungpook National University, Korea, in 1985 and 1987, respectively. He received the doctoral degree in Computer Science from the University of Massachusetts, USA, in 1999. He worked for ETRI between 1987 and 2000. Since September 2002, he is with the School of Electrical Engineering, Kookmin University, Seoul, Korea. He has organized several conferences such as ICUFN2009 and ICUFN2010. He is currently a member of the IEEE and KICS (Korea Information and Communications Society). He received the Young Science Award from the Korean Government (2003 to 2006). He had been the director of the Ubiquitous IT Convergence Center at Kookmin University since 2005. He has served as the executive director of KICS since 2006. In 2000, 2004, 2006, 2007 and 2009, he received the distinguished service medal from KICS. He has served as a founding chair of the KICS Technical Committee on Communication Networks in 2007 and 2008. His research interests include IMT-advanced, radio resource management, femtocell networks, Multi-screen convergence networks, and VLC WPANs.

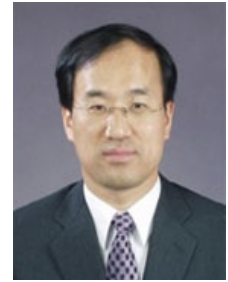

**Zygmunt J. Haas** received his B.Sc. in EE in 1979 and M.Sc. in EE in 1985. In 1988, he earned his Ph.D. from Stanford University and subsequently joined AT&T Bell Laboratories in the Network Research Department. There he pursued research on wireless communications, mobility management, fast protocols, optical networks, and optical switching. From September 1994 till July 1995, Dr. Haas worked for the AT&T Wireless Center of Excellence, where he investigated various aspects of wireless and mobile networking, concentrating on TCP/IP networks. In August 1995, he joined the faculty of the School of Electrical and Computer Engineering at Cornell University, where he is now a Professor.

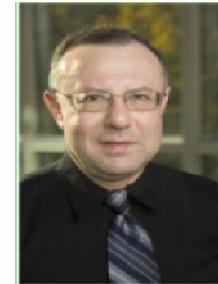

Dr. Haas is an author of numerous technical papers and holds eighteen patents in the fields of high-speed networking, wireless networks, and optical switching. He has organized several workshops, delivered numerous tutorials at major IEEE and ACM conferences, and has served as editor of a number of journals and magazines, including the IEEE Transactions on Networking, the IEEE Transactions on Wireless Communications, the IEEE Communications Magazine, the Springer "Wireless Networks" journal, the Elsevier "Ad Hoc Networks" journal, the "Journal of High Speed Networks," and the Wiley "Wireless Communications and Mobile Computing" journal. He has been a guest editor of IEEE JSAC issues on "Gigabit Networks," "Mobile Computing Networks," and "Ad-Hoc Networks." Dr. Haas served in the past as a Chair of the IEEE Technical Committee on Personal Communications (TCPC). He is an IEEE Fellow. His interests include: mobile and wireless communication and networks, biologically-inspired networks, and modeling of complex systems.